\documentclass[fleqn,usenatbib]{mnras}
\usepackage{newtxtext,newtxmath}
\usepackage[T1]{fontenc}

\DeclareRobustCommand{\VAN}[3]{#2}
\let\VANthebibliography\thebibliography
\def\thebibliography{\DeclareRobustCommand{\VAN}[3]{##3}\VANthebibliography}

\usepackage{graphicx}	
\usepackage{amsmath}	
\usepackage{pdflscape}

\title[A targeted search for FRB counterparts with KW]{A targeted search for FRB counterparts with Konus-\textit{Wind}}

\author[A. Ridnaia et al.]{
A. Ridnaia,$^{1}$\thanks{E-mail: ridnaia@mail.ioffe.ru}
D. Frederiks,$^{1}$
and D. Svinkin$^{1}$
\\
$^{1}$Ioffe institute, Politekhnicheskaya 26, St. Petersburg 194021, Russia}

\date{Accepted 2023 November 15. Received 2023 November 13; in original form 2023 September 29}

\pubyear{2023}

\begin{document}
\label{firstpage}
\pagerange{\pageref{firstpage}--\pageref{lastpage}}
\maketitle


\begin{abstract}

We present results of the search for hard X-ray/soft $\gamma$-ray emission in coincidence with publicly reported (via Transient Name Server, TNS~\footnote{Transient Name Server (TNS), http://www.wis-tns.org/}) fast radio bursts (FRBs). The search was carried out using continuous Konus-\textit{Wind} data with 2.944~s time resolution. We perform a targeted search for each individual burst from 581 FRBs, along with a stacking analysis of the bursts from 8 repeating sources in our sample and a separate stacking analysis of the bursts from the non-repeating FRBs. We find no significant associations in either case. We report upper bounds on the hard X-ray (20 - 1500~keV) flux assuming four spectral models, which generally describe spectra of short and long GRBs, magnetar giant flares, and the short burst, coincident with FRB~200428 from a Galactic magnetar. Depending on the spectral model, our upper bounds are in the range of $(0.1 - 2) \times10^{-6}$~erg cm$^{-2}$. For 18~FRBs with known distances we present upper bounds on the isotropic equivalent energy release and peak luminosity. For the nearest FRB 200120E, we derive the most stringent upper bounds of $E_{\text{iso}}\leq$2.0 $\times 10^{44}$~erg  and $L_{\text{iso}}\leq$1.2 $\times 10^{44}$~erg s$^{-1}$. Furthermore, we report lower bounds on radio-to-gamma-ray fluence ratio $E_{\text{radio}}/E_{\text{iso}} \geq 10^{-11}-10^{-9}$ and compare our results with previously reported searches and theoretical predictions for high-energy counterparts to FRBs. 

\end{abstract}




{   
 \renewcommand{\thefootnote}{\fnsymbol{footnote}}
 \footnotetext[2]{Transient Name Server (TNS), http://www.wis-tns.org/}
}

\section{Introduction}

Fast radio bursts (FRBs) are exceptionally bright ($\sim$Jy), short-duration~($\sim$ms) radio transients, discovered serendipitously in 2007 ~\citep{2007Sci...318..777L}. The dispersion measures (DM) of observed FRBs are well in excess of the expected Milky Way contribution, which implies they are originating from extragalactic distances (see, e.g. \cite{2019ARA&A..57..417C, 2022A&ARv..30....2P}, for a review). Over 600~unique sources have been reported thus far by different radio telescopes (see Table~\ref{tab:telescopes}), including 492 sources detected by the Canadian Hydrogen Intensity Mapping Experiment Fast Radio Burst (CHIME/FRB) Project~\citep{2021ApJS..257...59C}. Among them, only 18~\citep{2017Natur.541...58C, 2019Natur.572..352R,  2020ApJ...895L..37B, 2022AJ....163...69B} FRBs have been localized with enough (sub-arcsecond to arcsecond) precision to identify their host galaxies and redshifts, which confirms extragalactic origins and reveals a wide range of galaxy types and local environments surrounding the FRBs \citep{2020ApJ...903..152H}. More than half of these localizations have been provided by the Australian Square Kilometre Array Pathfinder (ASKAP; \citealt{2010PASA...27..272M}). While most FRBs are only seen once (``one-offs''), a small fraction ($\sim$~4~\%) of them have been found to produce multiple bursts (``repeaters'')~\citep{2016Natur.531..202S, 2021ApJS..257...59C, 2020ApJ...891L...6F, 2023ApJ...947...83A}. It remains an open question whether all FRBs repeat, and whether multiple progenitor populations of FRBs exist.

Until now, no clear physical picture of either the central-engine that produce a FRB or the mechanism by which the emission is generated has emerged. A wide range of models have been proposed, none of which is able to explain alone the variety of observed events (see~\citealt{2019PhR...821....1P} for a review). The most debated progenitor models include magnetars as their central-engines, with the FRB emission originating inside or outside of the magnetosphere~\citep{2010vaoa.conf..129P, 2020Natur.587...45Z, 2017MNRAS.468.2726K, 2014PhRvD..89j3009K, 2014MNRAS.442L...9L, 2019MNRAS.485.4091M, 2017ApJ...843L..26B, 2020ApJ...896..142B}. The recent discovery of a FRB-like event from the Galactic magnetar SGR~1935+2154 (FRB~200428; \citealt{2020Natur.587...59B,2020Natur.587...54C}) strongly suggests that at least some fraction of FRBs may originate from magnetars. The bright radio burst FRB~200428 was accompanied by the simultaneous emission of hard X-rays with properties similar to those of the short bursts typical of Galactic magnetars~\citep{2020ApJ...898L..29M, 2021NatAs...5..372R, 2021NatAs...5..378L, 2021NatAs...5..401T}, except for the peculiarly hard energy spectrum~\citep{2021NatAs...5..372R}. A couple more coincident radio and high energy events were detected from the same source~\citep{2022ATel15681....1D, 2022ATel15682....1W, 2022ATel15686....1F, 2022ATel15697....1M, 2022ATel15707....1H, 2022ATel15708....1L}, characterized by much fainter radio emission and longer duration than FRB~200428, and softer X-ray spectra typical of magnetar bursts.

To date, there is no other confirmed multi-wavelength or multi-messenger transient being associated with any FRB. The presence or absence of a simultaneous or delayed emission corresponding to FRBs in different wavebands would be essential to constrain the emission mechanisms and to identify the FRB progenitor(s). In the last years, many multi-wavelength searches for FRB counterparts have been carried out at all wavelengths with no confirmed results (see, e.g.,~\citealt{2021Univ....7...76N} for a review). In high-energy domain, a number of systematic searches has been made by using archival data of  \textit{Fermi}~(GBM, \citealt{2019A&A...631A..62M}; LAT, \citealt{2023arXiv230509428P}), \textit{INTEGRAL} (IBIS-ISGRI, \citealt{2021ApJ...921L...3M}), \textit{AstroSat}~(CZTI,  \citealt{2020ApJ...888...40A}), \textit{Insight}-HXMT (HE, \citealt{2020A&A...637A..69G}), \textit{AGILE} (MCAL, GRID; ~\citealt{2021ApJ...915..102V}), and data of multi-wavelength campaigns involving multiple instruments ~\citep{2019ApJ...879...40C, 2023A&A...676A..17T}. However, most of these studies were based on small FRB samples (less than 50~sources) or only focused on certain objects.

In this work, taking the advantages of the huge increase in the number of detected FRBs and continuous full-sky observations covering the entire current era of FRBs, performed by the Konus-\textit{Wind} $\gamma$-ray spectrometer (KW), we carry out a targeted search for possible hard X-ray/soft $\gamma$-ray counterparts to over 700 publicly reported bursts from repeating and non-repeating FRBs in KW archival data. The structure of this paper is the following. In Section~\ref{sec:analysis} we provide the FRB sample used in the search and briefly describe our search methodology and upper bound calculations. In Section~\ref{sec:results} we present our results, to then discuss it and provide our summary and future prospects in Section~\ref{sec:conclusions}.

\section{Data and analysis}
\label{sec:analysis}
\subsection{FRB sample}

\begin{table*}
\caption{List of radio telescope facilities with number of detected FRBs included in our sample.}
\label{tab:telescopes}
\centering
\begin{tabular}{lccc}
\hline
Facility & FRBs  & Frequency range (GHz) & Location \\
\hline
Canadian Hydrogen Intensity Mapping Experiment (CHIME)$^1$  & 566 & $0.4-0.8$ & Canada \\
Australian Square Kilometre Array Pathfinder (ASKAP)$^2$ & 47 & $0.7-1.8$ & Australia \\
National Astronomy and Ionosphere Center, NAIC (Arecibo)$^3$ & 12  & $0.1-11$ & Puerto Rico \\
Parkes Observatory$^4$ &33 & $0.6-26$ & Australia\\
Molonglo Observatory Synthesis Telescope (MOST)$^5$ &17 & $0.6-1.2$ & Australia \\
Robert C. Byrd Green Bank Telescope (GBT)$^6$ &15 & $0.3-110$ & USA\\
Deep Synoptic Array-110 (DSA-110)$^7$ &1 & $1.3-1.5$ &USA \\
Effelsberg 100-m Radio Telescope$^8$ &4 & $0.4-95$ & Germany\\
Five-hundred-meter Aperture Spherical radio Telescope (FAST)$^9$ &2 & $0.1-3$ &China \\
Giant Metrewave Radio Telescope (GMRT)$^{10}$ &4 & $0.2-1.4$ &India \\
Large Phased Array (LPA)$^{11}$ &10 &$0.109-0.111$ &Russia \\
Sardinia Radio Telescope (SRT)$^{12}$ &3 & $0.3-116$ & Italy \\
Very Large Array (VLA)$^{13}$ &4 & $0.1-50$ &USA \\
Westerbork Synthesis Radio Telescope (WSRT)$^{14}$  &3 &$0.1-8.3$ &Netherlands \\

\hline
\end{tabular}

{$^1$~\cite{2018ApJ...863...48C}
$^2$~\cite{2021PASA...38....9H}
$^3$~\cite{2006ApJ...637..446C}
$^4$~\cite{1996PASA...13..243S}
$^5$~\cite{1981PASA....4..156M, 1991AuJPh..44..729R}
$^6$~\cite{2009IEEEP..97.1382P}
$^7$~\cite{2023AAS...24123901R}
$^8$~\cite{2011JAHH...14....3W}
$^9$~\cite{2011IJMPD..20..989N}
$^{10}$~\cite{1991CSci...60...95S}
$^{11}$~\cite{2022muto.confE..43T}
$^{12}$~\cite{2017AA...608A..40P}
$^{13}$~\cite{2011ApJ...739L...1P}
$^{14}$~\cite{2009wska.confE..70O}}
\end{table*} 

\begin{figure}
	\includegraphics[width=\columnwidth]{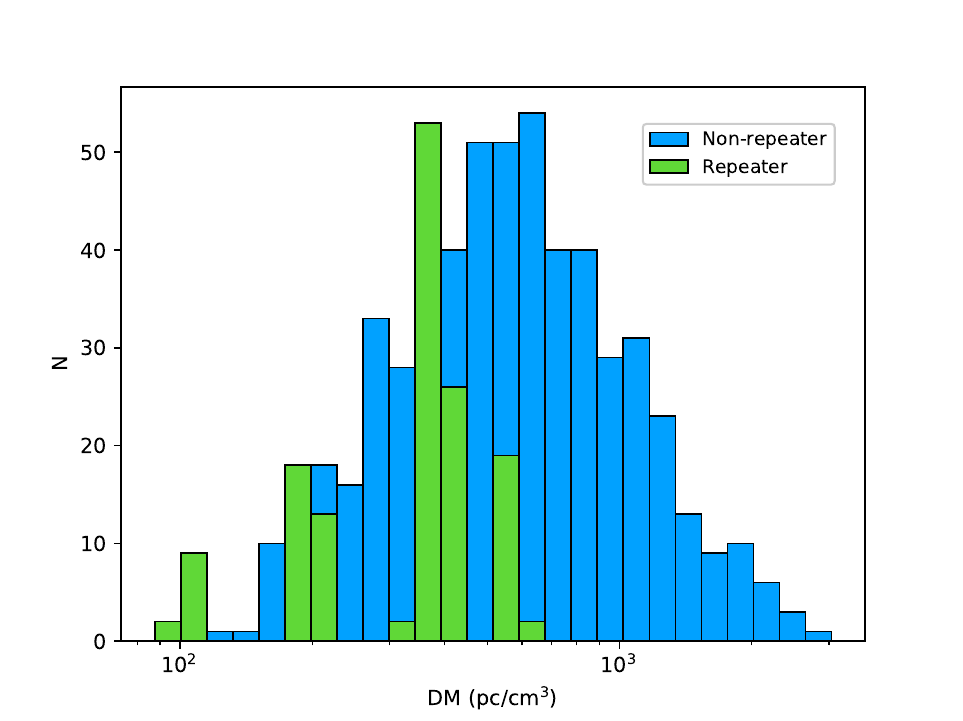}
   \caption{ Dispersion measure (DM) distribution of our selected sample of 581 FRBs: 573 ``one-off" 
  	 events and 148 bursts from 8 repeating sources. 18 FRB sources that have been associated with a 
   	host galaxies, have a luminosity distances range from 3.6 Mpc to 4 Gpc.}
   \label{samplestat}
\end{figure}

For our analysis we extract all publicly reported FRBs from TNS (799 events, accessed on 2022 April 27). Six events had to be discarded due to incomplete event information, such as FRB coordinates or burst time arrival, and 25 events due to gaps in the KW data at the time of interest. In addition, we decided to exclude 14 repeating sources, which have less than six bursts per source and have no accurate localization. Thus, the FRB sample used in our analysis consists of 721 events detected with 14 radio telescope facilities (see Table~\ref{tab:telescopes}) between 2001 January 25 and 2022 January 5. This includes 573 thus far one-off FRBs and 148 bursts from eight repeating sources: FRB~121102A, FRB~180814A, FRB~180916B, FRB~181030A, FRB~190303A, FRB~190711A, FRB~200120E, FRB~201124A. Full list of FRB events considered in the analysis and their measured parameters are available at the webpage~\footnote{http://www.ioffe.ru/LEA/FRB/}. Figure~\ref{samplestat} shows the dispersion measure distribution of the selected FRBs. To derive upper bounds on the radio-to-high-energy fluence ratio, we use fluence measurements from the first CHIME/FRB catalog~\citep{2021ApJS..257...59C}.

\subsection{Konus-\textit{Wind} analysis}

Konus-\textit{Wind} is a gamma-ray spectrometer which has been successfully operating since November 1994~\citep{Aptekar}. KW orbit is far from the Earth magnetosphere (since 2004 at distance of $\sim 5$ light seconds) that enables nearly uninterrupted observations of entire sky under very stable background. 
The continuous KW waiting-mode data consist of count rates in $\sim 20 - 80$~keV (G1), $\sim 80 - 320$~keV (G2), and $\sim 320 - 1300$~keV (G3) bands with temporal resolution of 2.944~s. These data are a valuable resource for various studies on hard X-ray/soft gamma-ray transients~\citep{2019JPhCS1400b2014K, Ridnaia:2020pau}. 

To search for FRB counterparts, we first estimate the burst arrival time $T_0$ at the KW position for each FRB. For this, we make two time corrections: a frequency-dependent time delay due to dispersion of the radio frequency with respect to soft $\gamma$-rays (infinite frequency) and a propagation time delay between KW and the telescope site. The combined corrections range from few milliseconds to hundreds of seconds, with a mean (median) value of 9.2 (4.9) s. We then search for significant ($ > 5 \sigma$) excess over background during the 400~s time interval centered on $T_0$. While the search interval length of 400~s is chosen arbitrarily, we were motivated by the discovery of the Galactic FRB~200428 accompanied by the simultaneous emission of hard X-rays and by theoretical predictions of very weak high-energy emission on time scales of (at most) minutes after the radio signal~\citep{2020MNRAS.498.1397L, 2019MNRAS.485.4091M}.
The search is performed in six energy channel combinations (G1, G2, G3, G1+G2, G2+G3 and  G1+G2+G3), on temporal scales from 2.944 s to 100 s, similar to~\cite{Svinkin2019}. The linear background approximation is estimated using two time intervals, before ($T_0$ - 1000 s, $T_0$ - 250 s) and after ($T_0$ +250 s, $T_0$ + 1000 s) the search interval. 

\subsubsection{Upper bound on the peak flux and fluence}
\label{sec:bounds}

In the case of non-detection of a significant counterpart in the KW data, we estimate upper bounds on its peak energy flux and energy fluence using four template spectral models, which represent typical short and long GRBs~(\citealt{Svinkin2016, 2017ApJ...850..161T}), huge initial pulses of magnetar giant 
flares (MGFs; ~\citealt{Svinkin:2021wcp}), and the Galactic SGR/FRB~200428 event~(\citealt{2021NatAs...5..372R}). These models  
are characterized by the Band function~\citep{1993ApJ...413..281B} or an exponentially cut off power law~(CPL), with the parameters listed in Table~\ref{tab:specTemplate}.

\begin{table}
\caption{The four source spectrum models used in upper bound calculations.}
\label{tab:specTemplate}
\centering
\begin{tabular}{ccccc}
\hline
Description & Model & \multicolumn{3}{c}{Parameters} \\
&            &$\alpha$ &$\beta$ & E$_\text{p}$ (keV)\\
\hline
Typical long GRB         &  Band   & -1.0       & -2.5  & 300\\
Typical short GRB        &  CPL    & -0.5       & ...      & 500\\
MGF (GRB~200415A)   &  CPL    & -0.6       & ...      & 1190 \\
SGR/FRB 200428        &  CPL    & -0.72     & ...      & 85   \\

\hline
\end{tabular}
\end{table}

\begin{figure}
	\includegraphics[width=\columnwidth]{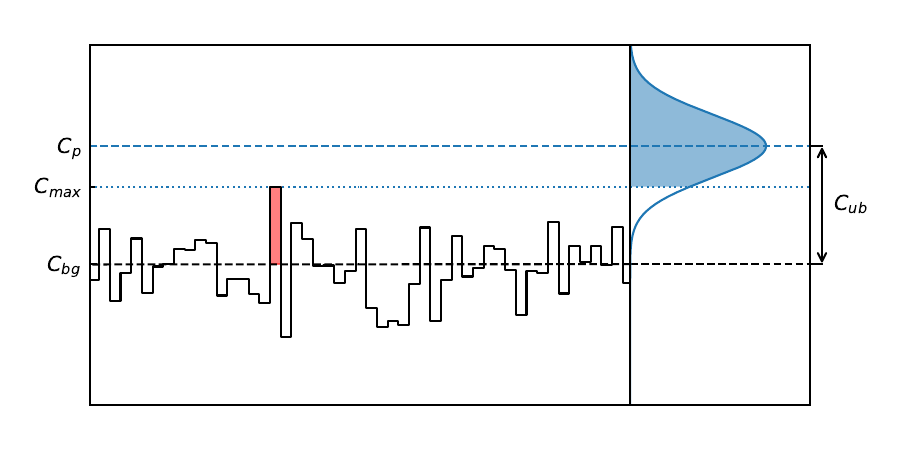}
   \caption{ Upper bound calculation. For the bin with the maximum observed count rate~$C_\mathrm{max}$ (red) we estimate the background count rate $C_\mathrm{bg}$ (the dashed line). We define upper bound on the source counts $C_\mathrm{ub}$ (corresponding to the 90~\% confidence level) so that the probability to observe $C > C_\mathrm{max}$, assuming that the counts have Gaussian distribution with  $\mu=\sigma^2=(C_\mathrm{bg} + C_\mathrm{ub})$, equals 0.9.}
   \label{illustration}
\end{figure}

In this work we use upper bound on the gamma-ray flux defined as the upper edge of a (frequentist) confidence interval for the flux of the 
source, according to~\cite{2010ApJ...719..900K}. To estimate an upper bound $C_\mathrm{ub}$ on the source counts in a particular KW light curve, measured in the energy band $\Delta E$, we use the bin with the maximum count rate, for which $C_\mathrm{max}$ is the observed number of counts, $C_\mathrm{bg}$ is the estimated number of background counts, and $\sigma_\mathrm{bg}$ is the error of the background estimation. We define $C_\mathrm{ub}$ (corresponding to the confidence level CL, hereafter 
CL=0.9) so that the probability to observe $C > C_\mathrm{max}$, assuming that the 
counts have Gaussian distribution with  $\mu=\sigma^2=(C_\mathrm{ub} + C_\mathrm{bg} + \sigma^2_\mathrm{bg})$, 
equals CL (see Figure~\ref{illustration}). We find that the last term $\sigma^2_\mathrm{bg}$ contribute less than a percent to the $\sigma^2$, and therefore can be omitted from our calculations.

The upper bound on the source counts then can be converted into a fluence (peak flux) upper 
bound in the standard energy range (20 -- 1500~keV) by using the count-to-energy conversion factor $k$ dependent on $\Delta E$, the 
template spectrum, the FRB sky location (the angle of incidence), and the corresponding KW detector 
response. The maximum value of $k C_\mathrm{ub}$ or $k C_\mathrm{ub}/2.944$~s is adopted as the upper bound on the corresponding short ($<2.944$~s) event energy fluence or the long event peak energy flux, respectively.

\subsubsection{Stacking analysis}

Current theories make widely varying predictions about FRB high-energy counterparts, with expected emission being faint (below the threshold sensitivity of the present telescopes) in most of the models (see e.g.~\citealt{2020ApJ...897..146C}). Assuming that parameters determining the hard X-ray/soft $\gamma$-ray emission have the same values for all the FRBs, we can employ the stacking analysis. The stacking analysis is a powerful technique that makes it possible to detect sources below the detection threshold. It brings down the statistical noise by combining the signal of many individually undetected sources. 

We perform a stacking analysis of the KW data by summing up the background subtracted count rates of the individual event light curves, centered on $T_0$ and then devided by the number of the summed events. To calculate upper bounds for the resulted light curve we use a similar procedure as described in \ref{sec:bounds}, except estimating an upper bound $R_\mathrm{ub}$ on rates instead of $C_\mathrm{ub}$ on counts. Two sets of upper bounds were computed: one based on the bin with the maximum count rate (assuming that all FRB events have the same large ($>3$~s) time delay between FRB and its high-energy counterpart) and the other, on the bin comprising $T_0$ (the non-delayed case).

We carry out a stacking analysis of the bursts from each repeating source in our sample and a separate stacking analysis of the bursts from the non-repeating FRBs.

\section{Results}
\label{sec:results}

\subsection{Candidate transients}
\label{sec:candidates}
\begin{figure*}
	\begin{minipage}{\columnwidth}
		\includegraphics[width=\columnwidth]{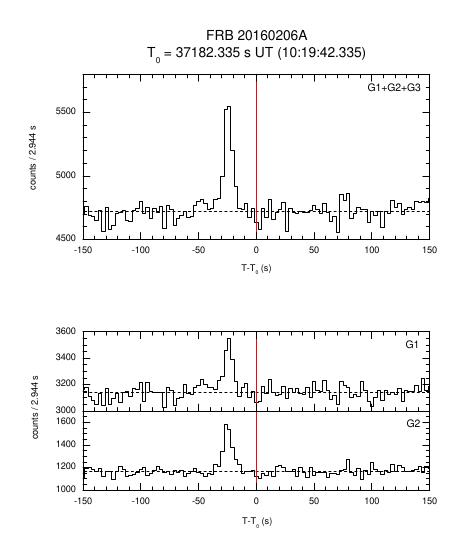}
	\end{minipage}
\hspace{0.5cm}
\begin{minipage}{\columnwidth}
	\includegraphics[width=\columnwidth]{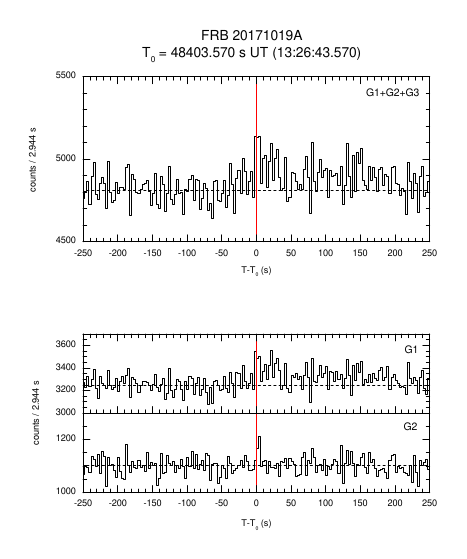}
\end{minipage}
   \caption{Candidate transient sources found for FRB~160206A ($12\sigma$ significance, left) and FRB~171019A ($5.3 \sigma$ significance, right). Arrival time of FRBs corrected for delays due to dispersion and propagation marked by red line.}
   \label{candidates}
\end{figure*}

Our search resulted in two candidate transient events, coincident in time with FRB~160206A and FRB~171019A (see Figure~\ref{candidates}). 
The first one turns out to be a GRB 160206B, which was also detected and localized by \textit{Fermi}-GBM (trigger 476446756/bn160206430).
The GRB localization is inconsistent with the FRB position, which lies far outside the 3~$\sigma$ GBM localization region. 

In the second case, a KW ecliptic latitude response~\citep{2022ApJS..259...34S} for the KW-detected transient is inconsistent with the position of FRB~171019A. Moreover, the FRB position is outside Earth-occulted part of the sky for \textit{Swift} and the source is located right at the edge of the BAT coded field of view, and so a FRB-related GRB might be captured by BAT as a count rate excess. We examine BAT data around the time of FRB~171019A and the KW transient~\footnote{https://heasarc.gsfc.nasa.gov/FTP/swift/data/obs/2017\_10/00780203000/bat/rate/} and find no significant count rate increase at 5~$\sigma$ level. This, together with the shape of the KW light curve and its detectors' response hints towards this transient being an accidentally coincident GRB.

Thus, we conclude that both candidate transients found in our search are physically unrelated to the FRBs. Based on the continuous KW observations between November 1994 and August 2017~\citep{2019JPhCS1400b2014K} we estimate an average KW GRB detection rate to $\sim 0.8$ GRB~per day. Assuming this rate, an expected number of GRBs detected by KW during the total exposure time of our search ($\sim3.25$ days) is  $2_{-2}^{+3}$ (95\% conf.), which is consistent with the number of the observed events.

\subsection{Upper Bounds}
\begin{figure}
	\includegraphics[width=\columnwidth]{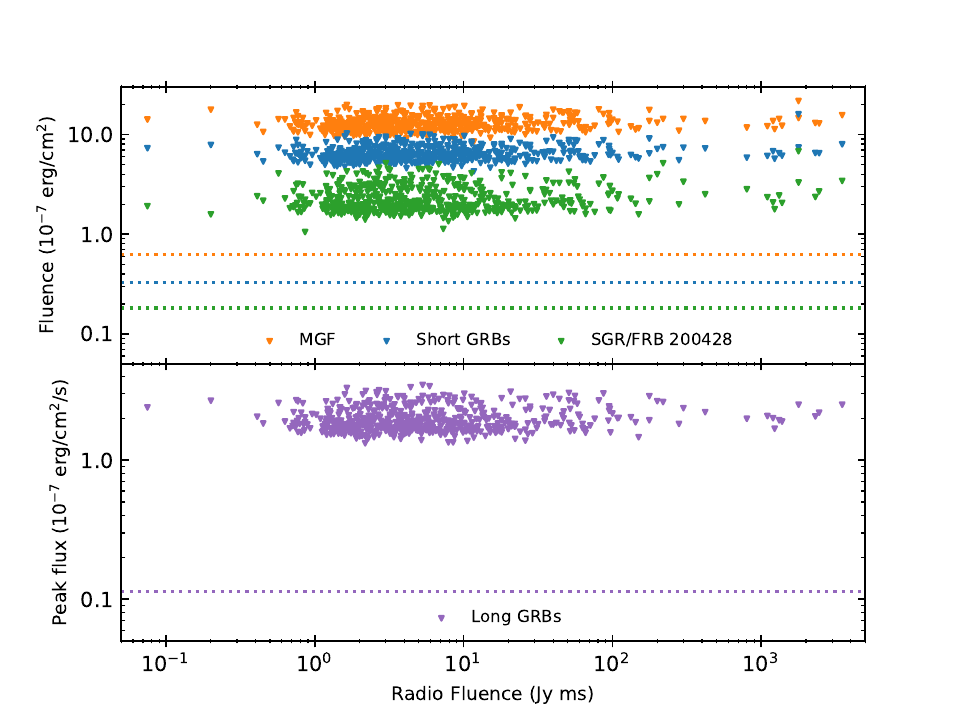}
   \caption{ KW upper bounds on the 20-1500 keV fluence (top panel) for a short bursts having typical KW short GRB/MGF/FRB~200428 spectrum and on peak flux (bottom panel) of long bursts derived for 581 FRBs from our sample. The dotted lines show the results of the stacking analysis of non-repeating FRBs.} 
   \label{res}
\end{figure}

Our search did not reveal any significant hard X-ray/soft $\gamma$-ray emission associated with the 721 FRB events reported through the TNS and detected between 2001 January 25 and 2022 January 5. Following the procedure of Section~\ref{sec:analysis} we have set upper bounds, that are presented in Table~\ref{tab:Ub}. 
The stacked data analysis allows us to set a factor of 20 (25 in the case of upper bounds based on the bin comprising $T_0$) on average more stringent than individual upper bounds.  Figure~\ref{res} summarises the results. 

\begin{table}
\caption{Upper bounds on the 20--1500~keV fluence (peak flux). }
\label{tab:Ub}
\centering
\begin{tabular}{lc}
\hline
Spectral template & Upper bounds range \\
& ($10^{-7}$~erg~cm$^{-2}$)\\
\hline
Typical long GRB$^{*}$         &  1 -- 4\\
Typical short GRB        &  5 -- 10\\
MGF, GRB~200415A   &  9 -- 20\\
SGR/FRB 200428        &  1 -- 7\\
\hline
\end{tabular}

{$^{*}$ For the long burst template, we provide upper bounds on the peak flux in units of $10^{-7}$  erg cm$^{-2}$ s$^{-1}$.}  
\end{table} 

For the FRBs with measured redshifts we estimate upper bounds on the total isotropic-equivalent energy release~$E_\text{iso}$ and peak luminosity~ $L_{\text{iso}}$ (see Table~\ref{tab:LisoEiso}). While we calculate these upper bounds for each of the four spectral templates, the bounds listed in Table~\ref{tab:LisoEiso} are given on $E_{\text{iso}}$ for short GRBs template and on $L_{\text{iso}}$ for long GRBs. On average, upper bounds for MGF spectral template results in a factor of two less stringent values and in a factor of two and a half more stringent values for SGR/FRB~200428 template. The upper bounds derived from the stacked data analysis are reported for the repeating FRBs (the upper bounds computed using $T_0$ bin are given in parenthesis).

\begin{table}
\caption{Upper bounds on the total isotropic equivalent energy release (short GRBs template) and peak luminosity (long GRBs template) for FRBs with measured redshifts. The upper bounds derived from the stacked data analysis are reported for the repeating FRBs.}
\label{tab:LisoEiso}
\centering
\begin{tabular}{lcrcc}
\hline
FRB & Repeating & Host redshift &  E$_{\text{iso}}$ &L$_{\text{iso}}$ \\
        &              &                      &  (10$^{49}$~erg) &(10$^{49}$~erg/s) \\
\hline
180924B & N &	0.3212$^{\text{b}}$          &21.76	&7.44\\
181112A &  N &	0.4755$^{\text{c}}$	    &38.44	&15.05\\
190102C & N &	0.2913$^{\text{b}}$ 	    &12.21	&4.32\\
190523A & N &	0.6600$^{\text{b}}$	    &66.64	&36.54\\
190608B & N &	0.1178$^{\text{b}}$ 	    &2.74	&1.07\\
190611B & N &	0.3778$^{\text{b}}$ 	    &28.17	&9.80\\
190614D & N&	0.60$^{\text{d}}$ 	    	    &65.34	&30.12\\
190714A & N&	0.2365$^{\text{b}}$	    &13.05	&5.23\\
191001A & N &	0.2340$^{\text{b}}$	    &8.40	&3.29\\
191228A & N &	0.2432$^{\text{b}}$	    &14.38	&5.67\\ 
200430A & N &	0.1608$^{\text{b}}$ 	    &3.92	&1.54\\
200906A & N &	0.3688$^{\text{b}}$	    &24.65	&10.61\\
121102A &	Y	 &     0.1927$^{\text{b}}$   &1.58 (0.38)	&0.64 (0.16)\\
180916B &	Y	 &     0.0337$^{\text{b}}$   &0.02 (0.001)	&67.19 (4.67) $\times10^{-4}$\\
181030A &	Y	 &     0.0039$^{\text{e}}$   &6.51(1.11)$\times10^{-4}$ & 1.96 (0.34)$\times 10^{-4}$\\
190711A &	Y	 &     0.5220$^{\text{b}}$  &28.31 (6.43)	& 14.59 (2.53)\\
200120E$^{\text{a}}$ &	Y	 &    (3.6 Mpc)$^{\text{f}}$  &1.96 (0.38)$\times 10^{-5}$	 & 1.18 (0.27) $\times 10^{-5}$\\
201124A &	Y	 &     0.0979$^{\text{b}}$    &0.32 (0.23)	&0.12 (0.08)\\ 
\hline
\end{tabular}

{$^{\text{a}}$ FRB source at a distance of 3.6~Mpc with a formally negative redshift
$^{\text{b}}$~\cite{2022AJ....163...69B}
$^{\text{c}}$~\cite{2019Sci...366..231P}
$^{\text{d}}$~\cite{2020ApJ...899..161L}
$^{\text{e}}$~\cite{2021ApJ...919L..24B}
$^{\text{f}}$~\cite{2022Natur.602..585K}}
\end{table} 

\begin{figure}
\includegraphics[width=\columnwidth]{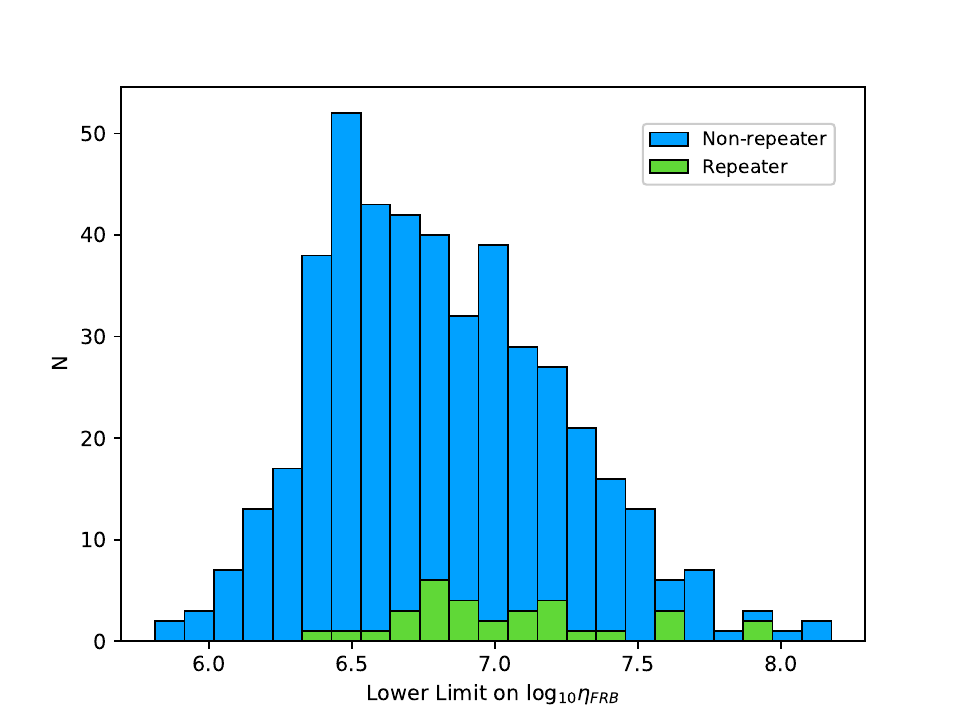}
\caption{Lower bounds on the radio-to-gamma-ray fluence ratio distribution of FRBs with radio fluences measured by CHIME/FRB. The bounds are in units of Jy ms erg$^{-1}$ cm$^2$.}
\label{fig:ratio}
\end{figure}

Using the derived fluence/peak flux upper bounds and the available radio fluences from the first CHIME/FRB catalog, we calculate the lower bounds on radio-to-gamma-ray fluence ratio $\eta_{\text{FRB}}$. The provided radio fluences are lower bounds due to the telescope sensitivity at the centre of the field of view is assumed~\citep{2021ApJS..257...59C}. We show the distribution of these ratios in Figure~\ref{fig:ratio} for the repeating and non-repeating FRBs from the joint TNS and CHIME/FRB sample. 

\section{Discussion and Conclusions}
\label{sec:conclusions}

Our results, derived with one of the largest FRB sample used so far, are consistent with that found from previous studies. \cite{2019ApJ...879...40C} searched for high-energy counterparts to 23~FRBs in GBM, LAT, and BAT data and found $\eta_{\text{FRB}} \geq 10^{5}-10^{7}$~Jy ms erg$^{-1}$ cm$^2$, which is comparable with the derived in this paper. A search for long-duration (1 to 200~s) $\gamma$-ray emission coincident with FRBs was carried out by~\cite{2019A&A...631A..62M} in cumulative GBM light curves. They obtained a deep upper limit $\eta_{\text{FRB}} > 10^{8}$~Jy ms erg$^{-1}$ cm$^2$. Both primary classes of FRB models (magnetospheric and maser-shock models) predict prompt high-energy counterparts and specify the ratio between the energy emitted by the counterpart and by the FRB itself~\citep{2019MNRAS.485.4091M, 2021MNRAS.508L..32C, 2021ApJ...919...89Y}. In order to compare our results with theoretical predictions, we have set limits on the radio-to-gamma-ray fluence ratio in dimensionless units. Assuming radio fluence and frequency bandwidth values from the literature (see Table~\ref{tab:radioloc}) for 12 non-repeating FRBs with known distances we found $E_{\text{radio}}/E_{\text{iso}} \geq 10^{-11}-10^{-9}$. Although the timescales and energy ranges are not identical to our analysis, this is consistent with the ratios obtained over different FRB samples with different instruments ($10^{-10} - 10^{-7}$,~\cite{2021Univ....7...76N}) and only approaches the ratios expected from theory ($10^{-6}$~\cite{2014MNRAS.442L...9L} to $10^{-5}$~\cite{2019MNRAS.485.4091M, 2021ApJ...919...89Y}). However, one should keep in mind that intrinsic fluence ratios may be significantly different from the observed ones due to beaming effects and that we need statistical limits on fluence ratio of lots of FRBs to constrain the models.

\begin{table}
\caption{Radio fluences for non-repeating FRBs with measured redshifts.}
\label{tab:radioloc}
\centering
\begin{tabular}{lcccc}
\hline
FRB & Fluence& Instrument & Cent. frequency & Bandwidth \\
        &      (Jy ms)           &                     &  (MHz) & (MHz)  \\
\hline
180924B$^\text{a}$ & $16\pm1$            & ASKAP    &1320	  	&  336\\
181112A$^\text{b}$ &  $26\pm 3$           & ASKAP	   &1295 	  	&336\\
190102C$^\text{c}$ & $14\pm1$            & ASKAP	   &1295 	  	&336\\
190523A$^\text{d}$ & $ \geq 280$         & DSA-10    &1411	  	&152.6\\
190608B$^\text{c}$ & $26\pm 4$           & ASKAP	   &1295 	  	&336\\
190611B$^\text{c}$ & $10\pm 2$           & ASKAP	   &1295    	&336\\
190614D$^\text{e}$ & $0.62\pm0.07$    &VLA	   &1400   		&1024\\
190714A$^\text{f}$ & $8\pm2$ 	         & ASKAP	   &1272.5 	&336\\
191001A$^\text{g}$ & $143\pm 15$        & ASKAP    & 920.5 	&336\\
191228A$^\text{h}$ & $40^{+50}_{-10}$ &ASKAP	   &1272.5 	&336\\ 
200430A$^\text{i}$ & $35\pm 4$ 		 &ASKAP     & 864.5		&336\\
200906A$^\text{j}$ & $59^{+25}_{-10}$ &ASKAP     &864.5 		&336\\
\hline
\end{tabular}
{$^{\text{a}}$~\cite{2019Sci...365..565B}
$^{\text{b}}$~\cite{2019Sci...366..231P}
$^{\text{c}}$~\cite{2020Natur.581..391M}
$^{\text{d}}$~\cite{2019Natur.572..352R}
$^{\text{e}}$~\cite{2020ApJ...899..161L}
$^{\text{f}}$~\cite{2019ATel12940....1B}
$^{\text{g}}$~\cite{2019ATel13166....1S}
$^{\text{h}}$~\cite{2019ATel13376....1S}
$^{\text{i}}$~\cite{2020ATel13694....1K}
$^{\text{j}}$~\cite{2022AJ....163...69B}}

\end{table}

Unfortunately, the extragalactic distances and the expected faintness of FRB counterparts put them below the detection thresholds of currently available telescopes, observing at frequencies above the radio band. The nearest and brightest FRBs are the most promising candidates for multi-wavelength observations that could strongly constrain FRB emission models as the models become more quantitative. At present, only two extragalactic FRBs are located in a relative proximity to us, i.e., FRB 181030A from a star-forming spiral galaxy NGC 3252 at  a distance of 20~Mpc~\citep{2021ApJ...919L..24B} and FRB 200120E from a globular cluster in M81 at 3.6~Mpc~\citep{2022Natur.602..585K}. 
From the stacked data analysis of nine bursts from FRB 181030A and six bursts from FRB 200120E in our sample, we derive the most stringent upper bounds on $E_{\text{iso}}\leq$ 6.5$\times 10^{45}$~erg and  $E_{\text{iso}}\leq$2.0 $\times 10^{44}$~erg for short bursts from FRB 181030A and  FRB 200120E, respectively.

Based on the bounds obtained from our observations, we can exclude GRBs with $E_{\text{iso}} \geq 7\times10^{50}$~erg, that are the majority of the observed by KW population ($\sim$ 97\%;~\citealt{2017ApJ...850..161T, 2021ApJ...908...83T}), as counterpart candidates of localized FRBs from our sample. A magnetar flare origin of FRBs is consistent with the derived bounds in terms of either gamma-ray energetics, or radio-to-gamma-ray fluence ratios. For almost all FRBs considered, we can not rule out the occurrence of an extragalactic MGF, with isotropic energy similar to or smaller than that of GRB~200415A ($E_{\text{iso}}\sim 1.3 \times 10^{46}$~erg) or GRB~051103 ($E_{\text{iso}}\sim 5.3 \times 10^{46}$~erg)~\citep{Svinkin:2021wcp}. MGFs with radio-to-gamma-ray fluence ratio similar to that of the 2004 giant flare from the Galactic magnetar SGR 1806-20 ($\eta_{\text{GF}}<10^{7}$~~Jy ms erg$^{-1}$ cm$^2$; ~\citealt{2016ApJ...827...59T}) are partly consistent with $\eta_{\text{FRB}} \geq 10^{6}-10^{8}$~Jy ms erg$^{-1}$ cm$^2$ derived in this paper. The SGR/FRB~200428 event having a radio-to-gamma-ray fluence ratio of $\sim 7 \times 10^{11}$~Jy ms erg$^{-1}$ cm$^2$ ($\sim 10^{-5}$ in dimensionless units)~\citep{2020Natur.587...54C, 2021NatAs...5..372R} is far above our lower bounds on radio-to-gamma-ray fluence ratio. The most stringent KW bounds are placed using the stacked data analisys of bursts from two nearest FRB repeaters, these bounds rule out MGFs, but do not rule out short magnetar bursts with the typical emitted energies below 10$^{42}$~erg.

Both detections and non-detections of FRB counterparts in multi-wavelength and multi-messenger search campaigns are of great importance. 
As in the case of other transient phenomena, for example, GRBs, collecting observational data at as wide as possible energy band
are crucial for progress towards our understanding of these enigmatic events. The search of high-energy FRB counterparts with KW is a work in progress. With the rapid growth of FRB population we will soon be able to study several more close-by sources, and hence, significantly tighten bounds reported here.

\section*{Acknowledgements}

The authors acknowledge support from the Russian Science Foundation (RNF) grant 21-12-00250.

\section*{Data Availability}
KW data underlying this article will be shared from the corresponding author upon reasonable request.
All the information of the FRB events in our sample together with the upper bounds on their high energy emission obtained in our analysis
are available at the webpage http://www.ioffe.ru/LEA/FRB/. An example of tables provided online is given in the appendix.



\bibliographystyle{mnras}
\bibliography{example} 




\appendix

\section{FRB data tables}

The appendix gives an example of the FRB data tables, which can be found in full, machine-readable format online.

\begin{landscape}
 \begin{table}
 \scriptsize
 \caption{Properties of the bursts from FRB 20200120E}
  \label{tab:landscape}
  \begin{tabular}{cccccccccccccccccccc}
    \hline
FRB & Day & Arrival Time & RA & Dec& RA$_{\text{err}}$ & Dec$_{\text{err}}$ & DM  & DM$_\text{err}$ & Facility & Freq.& bandwidth & RFluence & PDelay & DDelay & KW Time & UB$_\text{SGRB}$ & UB$_\text{LGRB}$ & UB$_\text{MGF}$ & UB$_\text{SGR}$\\
                  &  (yyyymmdd)      & UTC          & deg  &deg  &  & &pc cm$^{-3}$ & & &MHz & MHz&Jy ms & s &s & UTC& erg cm$^{-2}$ & erg cm$^{-2}$ s$^{-1}$ & erg cm$^{-2}$  & erg cm$^{-2}$\\   
    \hline
20200120E & 20200120 & 35856.006 & 146.25 & 68.77 & 0.533 & 1.567 & 88.96 & 1.62 & CHIME & 600 & 400 & - & 3.272 & 1.0253 & 35858.253 & 5.44E-07 & 1.68E-07 & 1.08E-06 & 1.87E-07\\
20200718A & 20200718 & 79951.867 & 149.1 & 68.79 & 0.600 & 1.433 & 88.96 & 1.62 & CHIME & 600 & 400 & - & -3.405 & 1.0253 & 79947.437 & 5.04E-07 & 1.59E-07 & 9.94E-07 & 1.78E-07\\
20201129A & 20201129 & 48689.858 & 149.43 & 68.77 & 0.483 & 1.283 & 87.75 & 0.4 & CHIME & 600 & 400 & 2.4 & 2.479 & 1.0113 & 48691.325 & 6.22E-07 & 1.94E-07 & 1.26E-06 & 2.38E-07\\
20210423G & 20210423 & 13714.726 & 149.43 & 68.77 & - & - & 87.75 & 0.4 & CHIME & 600 & 400 & - & 0.547 & 1.0113 & 13714.261 & 6.15E-07 & 1.65E-07 & 1.22E-06 & 1.71E-07\\
20210430G & 20210430 & 11942.401 & 149.43 & 68.77 & - & - & 87.75 & 0.4 & CHIME & 600 & 400 & - & -0.077 & 1.0113 & 11941.313 & 4.63E-07 & 1.55E-07 & 9.17E-07 & 1.72E-07\\
20210823C & 20210823 & 71036.557 & 149.43 & 68.77 & - & - & 87.75 & 0.4 & CHIME & 600 & 400 & 1.4 & -3.499 & 1.0113 & 71032.047 & 5.52E-07 & 1.70E-07 & 1.14E-06 & 1.86E-07\\
    \hline
  \end{tabular}
 \end{table}
\end{landscape}


\label{lastpage}
\end{document}